\documentclass[aps,prb,twocolumn,showpacs,amsfonts,superscriptaddress]{revtex4}

\usepackage{epsfig}
\usepackage{amsmath}
\usepackage{amssymb}
\usepackage{bm}

\newcommand{\beq}{\begin{equation}}
\newcommand{\eeq}{\end{equation}}
\newcommand{\beqarray}{\begin{eqnarray}}
\newcommand{\eeqarray}{\end{eqnarray}}

\newcommand{\eq}[1]{Eq.~(\ref{#1})} 
\newcommand{\fig}[1]{Fig.~(\ref{#1})} 
\newcommand{\Ref}[1]{Ref.~\onlinecite{#1}} 

\begin{document}

\title{Spontaneous spin current due to triplet superconductor--ferromagnet interfaces} 
\author{P. M. R. Brydon}
\email{brydon@theory.phy.tu-dresden.de}
\affiliation{Institut f\"{u}r Theoretische Physik, Technische Universit\"{a}t
  Dresden, 01062 Dresden, Germany}

\date{\today}

\begin{abstract}
We examine the appearance of a spontaneous bulk spin current in a triplet
superconductor 
in contact with a metallic ferromagnet. The spin
current results from the spin-flip of Cooper pairs upon reflection from the
interface with the ferromagnet, and is shown to display strong similarities
to the spontaneous charge current in a Josephson junction. We express the
spin current 
in terms of the Andreev reflection coefficients, which are derived
by the construction of the quasiclassical scattering wavefunctions. The
dependence of the spin current upon a number of parameters is investigated, in
particular the orientation of the magnetic moment of the
ferromagnet, the exchange splitting, the temperature, and the orbital pairing
state of the triplet superconductor. 
\end{abstract}


\maketitle

\section{Introduction}

The interface between a singlet superconductor (SSC) and a ferromagnet (FM)
is an ideal setting to explore the antagonistic relationship between these
two phases.~\cite{Buzdin2005,Bergeret2005} It is now more than 25 years
since the first theoretical investigations,~\cite{Buzdin1982} but the physics
of SSC-FM interfaces continues to fascinate and
surprise.~\cite{Buzdin2005,Bergeret2005,Bergeret2001,Demler1997,Zhu2000,spinactivebcs,EschrigHF,Halterman2002,Halterman2007,Grein2009,Linder2007,Kashiwaya1999,Linder2009}
A key feature of such systems is the existence of an unconventional
proximity
effect:~\cite{Bergeret2005,Bergeret2001,EschrigHF,Halterman2002,Halterman2007,Grein2009,Linder2009}
in contrast to the spin-singlet pairing 
state of the bulk SSC, the spin-splitting of the Fermi surface induces
spin-triplet {\it correlations} in the FM. Although this effect
occurs at all metallic FM interfaces, the character of the induced triplet
pairing correlations is determined by the strength of the exchange-splitting
in the 
FM,~\cite{Bergeret2005,Bergeret2001,EschrigHF,Grein2009} whether the system is
in the ballistic or the diffusive 
limits,~\cite{Bergeret2005,EschrigHF,Halterman2007,Linder2009} and also the
particular geometry of the
heterostructure.~\cite{Bergeret2005,Halterman2007,Linder2009}   

The unconventional proximity effect at SSC-FM interfaces clearly evidences an
intimate connection between magnetism and spin-triplet pairing. Such devices
can provide only a limited understanding of this interplay, however, as the FM
determines the 
triplet pairing correlations. To be able to control the spin triplet pairing
independently, it would be necessary to replace the SSC  
with a triplet superconductor (TSC). Since the discovery of triplet
superconductivity in Sr$_2$RuO$_4$,~\cite{Maeno1994,Mackenzie2003} there has
been steadily growing interest in the properties of TSC
heterostructures.~\cite{Matsumoto1999,Barash2001,Kwon2004,Asano2004,Rashedi,Asano,Kastening2006,Tanuma2006,Norway,BrydonTFT2008,Brydonspin2008,BrydonCFC2009,Brydonlett2009,Kuboki2004,Cuoco2008,Yokoyama2007a,Yokoyama2007b,Tanaka2007,Hirai}
Despite the likely intimate connection between the two phases, the
study of devices combining TSCs and FMs is still in its 
infancy.~\cite{Kuboki2004,Cuoco2008,BrydonTFT2008,Brydonspin2008,BrydonCFC2009,Brydonlett2009,Yokoyama2007a,Yokoyama2007b,Hirai} 
Even so, several exotic effects have already been predicted, such as
a $0$-$\pi$ transition in a TSC-FM-TSC Josephson junction caused by the
mis-alignment of the vector order parameters (the so-called ${\bf d}$-vectors)
of the TSCs with the moment ${\bf M}$ of the FM tunneling
barrier.~\cite{Kastening2006,BrydonTFT2008,Brydonlett2009}  

The origin of this unconventional behaviour is the coupling of the FM moment
to the spin of the triplet Cooper pair.~\cite{Brydonlett2009} More generally,
the extra degree of freedom provided by the Cooper pair spin 
is responsible for novel spin transport properties of TSC
heterostructures. For example, 
a number of authors have demonstrated that a Josephson {\it
  spin current} flows between two TSCs when their ${\bf d}$ vectors are
misaligned.~\cite{Rashedi,Asano,Norway} It has
recently been established that a spin current may also be produced by the
inclusion of a FM tunneling barrier in a TSC Josephson
junction.~\cite{BrydonTFT2008,Brydonspin2008,BrydonCFC2009,Brydonlett2009}
Two basic mechanisms have been identified: the barrier can act as a
spin-filter, preferentially allowing the tunneling of one spin species of
Cooper pair over
the other; alternatively, the barrier moment can flip the spin of a tunneling
Cooper pair, which then acquires an extra spin-dependent phase.
However produced, the tunneling spin currents are always dependent upon the
phase difference between the TSC condensates on either side of the junction.

A bulk {\it phase-independent} contribution to the spin current in a
TSC-FM-TSC Josephson junction
was predicted in~\Ref{Brydonlett2009}, and subsequently also
identified in a SSC-FM 
heterostructure in~\Ref{Grein2009}. The origin of this
spin current was shown to be the spin-dependent phase shift acquired by the
flipping of a triplet
Cooper pair's spin upon reflection at a $\delta$-function-thin FM barrier. A
number of 
properties were deduced: the spin current is polarized along the direction
${\bf d}\times{\bf M}$; the current in the tunneling limit is
$\propto\sin(2\alpha)$, 
where $\alpha$ is the angle between ${\bf d}$ and ${\bf M}$; and the sign of
the current displays a pronounced dependence upon the orbital structure
of the bulk TSC, due to the orbital-dependent phase shift experienced by the
reflected Cooper pairs. Most remarkable is that although this spin current is
carried by Cooper pairs, and hence is indistinguishable from the tunneling
Josephson spin current, it is independent of the material on the other side
of the FM barrier.  

\begin{figure}
  \includegraphics[width=\columnwidth]{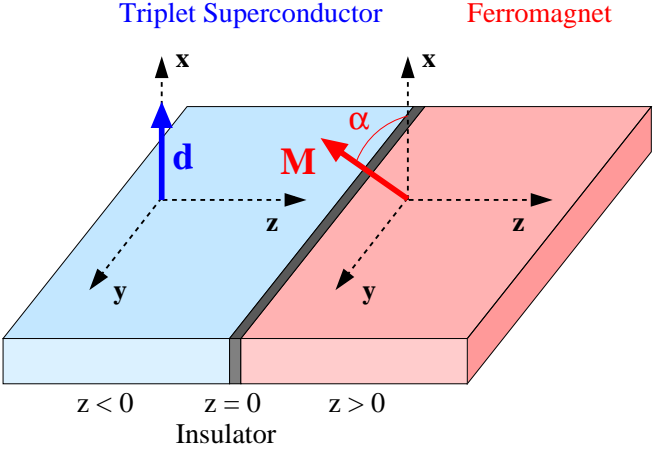}
  \caption{\label{device}(color online) Schematic representation of the
    device studied in this work. We consider a junction between a bulk triplet
    superconductor and a bulk ferromagnet, separated by a thin insulating
    tunneling barrier. The ${\bf d}$-vector of the TSC defines the $x$-axis,
    while the moment ${\bf M}$ of the FM lies in
    the $x$-$y$ plane at an angle $\alpha$ to the $x$-axis.}
\end{figure}

It is a natural question to ask if this effect also occurs at the interface
between a bulk TSC and FM. In studying such interfaces, most authors have
only addressed the case ${\bf d}\parallel{\bf M}$ when the spin current
described above is not expected to
occur.~\cite{Kuboki2004,Yokoyama2007b,Cuoco2008} Although Hirai and co-workers
considered arbitrary orientation of ${\bf d}$ and ${\bf M}$ in their study of
the TSC-FM interface, they did not
examine the spin transport properties of the device.~\cite{Hirai} It is
therefore the purpose of this 
paper to examine the occurrence of a spontaneous spin current in a TSC-FM
junction for arbitrary alignment of the TSC and FM vector order
parameters. Using a quasiclassical technique, we demonstrate the existence of
a spin current with the same dependence upon the bulk TSC orbital structure
and the relative misalignment of ${\bf d}$ and ${\bf M}$ as predicted
in~\Ref{Brydonlett2009}. The spin current is also found to display a strong
dependence upon the exchange-splitting of the FM, which is explained due to
the angular-dependence of the spin-flip reflection
probability. We discuss similarities and differences to the usual Josephson
effect, in particular arguing that low-temperature anomalies in the spin current
imply a role for zero energy states at the interface in the
transport. 

\section{Theoretical Formulation}

A schematic diagram of the device studied here is shown in~\fig{device}. It
consists of a bulk TSC and FM, separated by a thin insulating barrier at
$z=0$, which we approximate by a 
delta function of height $U$. Both materials are assumed to be in the clean
limit. The Bogoliubov-de Gennes (BdG) equation describing
the quasiparticle states with energy $E$ is written in Nambu-spin space as
\beq
\left(\begin{array}{cc}
\hat{H}_{0}({\bf r}) & \hat{\Delta}({\bf r}) \\
\hat{\Delta}^{\dagger}({\bf r}) & -\hat{H}^{T}_{0}({\bf r})
\end{array}\right)\Psi({\bf r}) = E\Psi({\bf r}) \label{eq:BdG}
\eeq
where the caret indicates a $2\times2$ matrix in spin-space.
The non-interacting Hamiltonian is
\beq
\hat{H}_{0}({\bf r}) = \left[-\frac{\hbar^2\pmb{\nabla}^2}{2m} +
  U\delta(z)\right]\hat{\mathbf{1}} - g\mu_{B}\hat{\pmb{\sigma}}\cdot{\bf M}\Theta(z)  
\eeq
In the interests of simplicity, we will assume that the effective mass $m$ is
the same in the TSC and the FM.~\cite{simplify} The moment of the FM
is parameterized  
as ${\bf M} = M[\sin(\beta)\cos(\alpha){\bf e}_{x} +
  \sin(\beta)\sin(\alpha){\bf e}_y + \cos(\beta){\bf e}_z]$. Since we do not
include spin-orbit coupling, we can assume without loss
of generality that $\beta=\pi/2$, i.e. the moment lies in the
$x$-$y$ plane. Other spin-orientations can be
obtained by appropriate rotation of the system about the $x$-axis in spin
space, which leaves the spin state of the TSC unchanged. 

\begin{figure}
  \includegraphics[width=\columnwidth]{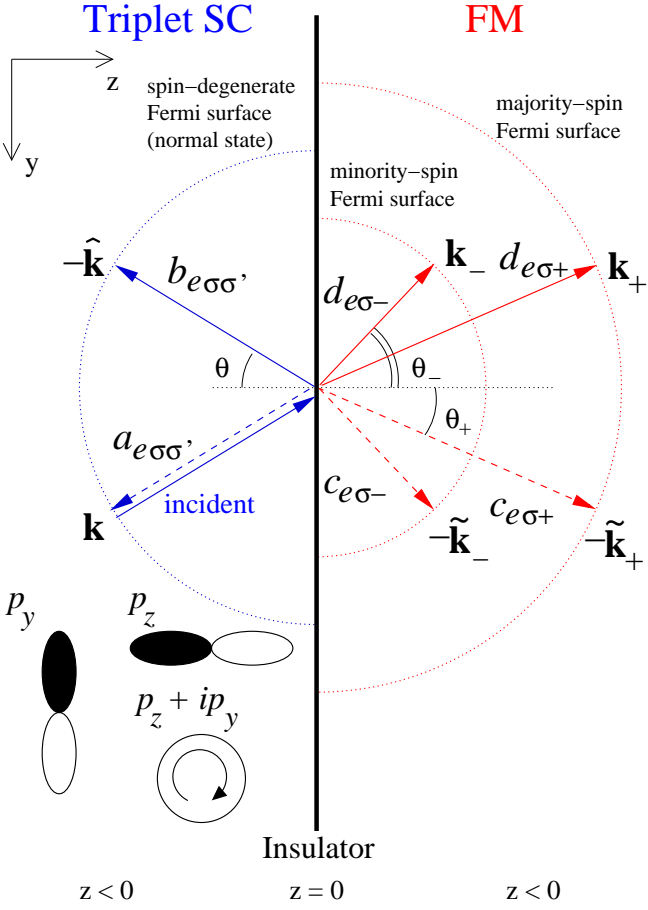}
  \caption{\label{wavefunction} (color online) Schematic representation of the
  wavefunction $\Psi_{e\sigma}$ for a spin-$\sigma$ electron-like
  quasiparticle with wavevector ${\bf k}$ incident upon the FM from the
  TSC. Note the different lengths of the wavevectors in the FM due to the
  exchange-splitting of the Fermi surface. In the bottom
  left corner, we show the orbital arrangement of our three choices of pairing
  symmetry. In the $p_y$ and $p_z$ cases, the black and white lobes of the
  $p$-wave orbital indicate opposite signs; for the $p_{z}+ip_y$ state, the
  gap magnitude is constant and the arrow indicates the direction of
  increasing phase.}    
\end{figure}

The gap matrix in~\eq{eq:BdG} is
$\hat{\Delta}({\bf r}) = i[\hat{\pmb{\sigma}}\cdot{\bf d}({\bf
    r})]\hat{\sigma}_y$ where ${\bf d}({\bf r})$ is the vector order parameter
of the TSC. We will restrict ourselves here to equal-spin-pairing unitary
states, for which ${\bf d}({\bf r}) = \Delta({\bf 
  r})\Theta(-z){\bf e}_x$ is a 
suitable choice. In such a state, the 
triplet Cooper pairs have $z$-component of spin $S_{z}=\pm\hbar$ but the
condensate has no net spin. The
magnitude of the gap is assumed to be constant throughout the TSC. Due to the
triplet spin state, the gap must reverse sign across the Fermi surface,
implying an odd-parity orbital wavefunction. We
will mostly be concerned with three different orbital pairing states:
$p_{y}$-wave, $\Delta_{\bf k}=\Delta(T)k_{y}/k_{F}$; $p_{z}$-wave, $\Delta_{\bf
  k}=\Delta(T)k_{z}/k_{F}$; and a chiral $p_{z}+ip_{y}$-wave state
$\Delta_{\bf k} = \Delta(T)[k_{z}+ik_{y}]/k_{F}$. These different gaps are
illustrated in~\fig{wavefunction}. The gap magnitude
$\Delta(T)$ displays weak-coupling temperature dependence, with $T=0$ value
$\Delta_{0}$. 

Both the TSC and FM are assumed to have circular Fermi surfaces. For
the sake of clarity, we assume the TSC and FM to have identical Fermi
energies $E_F$.~\cite{simplify} In 
the TSC, the Fermi surface in the normal state is spin-degenerate with radius
$k_{F}=\sqrt{2mE_{F}/\hbar^2}$. If the exchange-splitting of the bands is less
than the Fermi energy in the FM, we have a majority spin (aligned parallel to
${\bf M}$, $s=+$) and a minority spin (aligned anti-parallel to ${\bf M}$,
$s=-$) Fermi surface of radius
\beq
k^{s}_{F} = \sqrt{\frac{2m}{\hbar^2}(E_{F}+sg\mu_BM)} = k_{F}\sqrt{1+s\lambda}
\eeq
where 
\beq
\lambda=g\mu_{B}M/E_{F}
\eeq
is the ratio of the exchange-splitting to the 
Fermi energy. The minority spin Fermi surface disappears when the
exchange-splitting exceeds $E_{F}$, i.e. the FM is a half-metal. 

Solving the BdG equations~\eq{eq:BdG}, 
we construct the wavefunction for a spin-$\sigma$ electron-like quasiparticle
with wavevector ${\bf k}$ incident upon the FM layer from the
TSC using the Ansatz 
\beqarray
\lefteqn{\Psi_{e\sigma}({\bf r})}\notag \\ & = & \Theta(-z)\left\{\Phi^{TSC}_{{\bf k},e,\sigma}e^{i{\bf
    k}\cdot{\bf r}} +
\sum_{\sigma'=\uparrow,\downarrow}\left[a_{e\sigma\sigma'}\Phi^{TSC}_{{\bf
      k},h,\sigma'}e^{i{\bf k}\cdot{\bf r}}\right. \right. \notag \\
&& \left. \left. + b_{e\sigma\sigma'}\Phi^{TSC}_{-\hat{\bf
      k},e,\sigma'}e^{-i\hat{\bf k}\cdot{\bf r}}\right]\right\} \notag \\
&& +
\Theta(z)\sum_{s=\pm}\left\{c_{e\sigma{s}}\Phi^{FM}_{h,s}e^{-i\widetilde{\bf 
    k}_{s}\cdot{\bf r}} + d_{e\sigma{s}}\Phi^{FM}_{e,s}e^{i{\bf 
    k}_{s}\cdot{\bf r}}\right\} \label{eq:Psiesigma}
\eeqarray
A schematic representation of the wavefunction $\Psi_{e\sigma}$ is shown
in~\fig{wavefunction}. For $z<0$, the Ansatz~\eq{eq:Psiesigma} describes an
Andreev-reflected spin-$\sigma'$ hole-like 
quasiparticle with wavevector ${\bf k}$ and reflection
probability amplitude 
$a_{e\sigma\sigma'}$, and a spin-$\sigma'$ electron-like quasiparticle
undergoing specular 
reflection with wavevector $-\hat{\bf k} =(k_x,k_y,-k_z)$ and probability
amplitude $b_{e\sigma\sigma'}$. Note that we make the standard
``quasiclassical'' assumption that $E\ll{E_{F}}$, and so the magnitude of the
wavevectors for the electron-like and hole-like quasiparticles are
approximated to be identical.~\cite{Andreev1964,Kashiwaya2000}  
For $z>0$, the
transmission probability amplitudes for hole and electron quasiparticles with
spin projection $s=\pm$ along the direction of the ferromagnetic moment are
$c_{e\sigma{s}}$ and $d_{e\sigma{s}}$ respectively. 
As shown in~\fig{wavefunction}, due to the spin-splitting of the FM Fermi
surface, the trajectories of transmitted $s=+$ and $s=-$ quasiparticles are
not coincident. Since translational invariance is satisfied along the $x$ and
$y$ directions, the component of
the wavevector parallel to the interface is preserved during the
scattering. We hence have $k_{F}\sin(\theta) = k_{F}^{s}\sin(\theta_s)$. For
$\sin(\theta)<k_{F}^{s}/k_{F}$ propagating solutions for both spin
polarizations exist, and the $z$-component of 
${\bf k}_{s}$ and $\widetilde{\bf k}_{s}$ is then $k_{s,z} =
k_{F}\sqrt{\cos^{2}(\theta)+s\lambda}$. Since
$k_{F}^{-}<k_{F}<k^{+}_F$, there is however a critical angle
$\theta_c=\arcsin(k_{F}^{-}/k_{F})$ such that for $\theta_{c}<|\theta|<\pi/2$
the $z$-component of the wavevector of the transmitted $s=-$ quasiparticles is
purely imaginary. The resulting evanescent wave is exponentially suppressed
on an inverse length scale $\kappa=k_{F}\sqrt{\lambda - \cos^2(\theta)}$.

The wavefunction Ansatz~\eq{eq:Psiesigma} is expressed in terms of the spinors
for the bulk TSC and FM phases. 
In the FM we have the spinors for electrons and holes with spin $s=\pm$
\beqarray
\Phi^{FM}_{e,s} & = &
\left(se^{-i\alpha}/\sqrt{2},1/\sqrt{2},0,0\right)^T
\\
\Phi^{FM}_{h,s} & = &
\left(0,0,se^{i\alpha}/\sqrt{2},1/\sqrt{2}\right)^T
\eeqarray
For the TSC, the spinors $\Phi^{TSC}_{{\bf k},e(h),\sigma}$ for an 
electron-like (hole-like) quasiparticle with spin
$\sigma$ and wavevector ${\bf k}$ are
\beqarray
\Phi^{TSC}_{{\bf k},e,\uparrow} & = & \left(s_{\bf k}u_{\bf k},0,-v_{\bf k},0\right)^{T} \\
\Phi^{TSC}_{{\bf k},h,\uparrow} & = & \left(s_{\bf k}v_{\bf k},0,-u_{\bf k},0\right)^{T} \\
\Phi^{TSC}_{{\bf k},e,\downarrow} & = & \left(0,s_{\bf k}u_{\bf k},0,v_{\bf k}\right)^{T} \\
\Phi^{TSC}_{{\bf k},h,\downarrow} & = & \left(0,s_{\bf k}v_{\bf k},0,u_{\bf k}\right)^{T}
\eeqarray
where $u_{\bf k} = \sqrt{(E+\Omega_{\bf k})/2E}$, $v_{\bf k} =
\sqrt{(E-\Omega_{\bf k})/2E}$, $\Omega_{\bf k} = \sqrt{E^2 - |\Delta_{\bf
    k}|^2}$ and $s_{\bf k} = \Delta_{\bf k}/|\Delta_{\bf k}|$. 

The probability amplitudes in~\eq{eq:Psiesigma} are determined by the boundary
conditions obeyed by the wavefunction at the TSC-FM interface. In particular,
we require that the wavefunction is continuous at the interface, i.e. 
\beq
\Psi_{e\sigma}({\bf r})|_{z=0^{-}} = \Psi_{e\sigma}({\bf
  r})|_{z=0^{+}} \label{eq:continuity} 
\eeq
and also that the first derivative of the wavefunction obeys the condition
\beq
\left.\frac{\partial\Psi_{e\sigma}({\bf r})}{\partial{z}}\right|_{z=0^{+}} -
\left.\frac{\partial\Psi_{e\sigma}({\bf r})}{\partial{z}}\right|_{z=0^{-}} =
Zk_{F}\Psi_{e\sigma}({\bf r})|_{z=0} \label{eq:firstderiv}
\eeq
where $Z=2mU/\hbar^2k_{F}$ is a dimensionless constant characterizing the
height of the insulating barrier. These conditions yield eight coupled
equations for the probability amplitudes. In general, the
probability amplitudes are functions of the incident wavevector ${\bf k}$
and the quasiparticle energy $E$.

It is straight-forward to
modify the Ansatz~\eq{eq:Psiesigma} for the wavefunctions
$\Psi_{h\sigma}({\bf r})$ describing the scattering of a hole-like
quasiparticle incident upon the FM from the TSC. The probability
amplitudes $a_{h\sigma\sigma'}$, $b_{h\sigma\sigma'}$ etc. for this case are
determined by applying 
the same boundary conditions as for $\Psi_{e\sigma}({\bf r})$.

The currents in the TSC may be evaluated using the generalization of the
Furusaki-Tsukuda formula to triplet pairing.~\cite{FT1991,Asano} For the
currents flowing perpendicular to the interface,  
we hence express the charge current $I_{C}$ and the
$z$- and $y$-components 
of the spin current, $I_{S,z}$ and $I_{S,y}$ respectively, in terms of the
Andreev reflection probability amplitudes $a_{e(h)\sigma\sigma'}$ with energy
argument analytically continued to $i\omega_{n}$:
\beqarray
\lefteqn{I_{C}} \notag \\ & = & \frac{e}{2\hbar}\int_{|{\bf k}|=k_{F}}dk_{z}dk_{y}\Theta(k_{z}) \frac{k_{z}}{k_{F}}\frac{1}{\beta\hbar}\sum_{n}\sum_{\sigma}
\notag \\
& & \times\left\{\frac{|\Delta_{\bf k}|}{\Omega_{n,{\bf
    k}}}a_{e\sigma\sigma}({\bf k},i\omega_n) - 
\frac{|\Delta_{-\hat{\bf k}}|}{\Omega_{n,-\hat{\bf
    k}}}a_{h\sigma\sigma}({\bf k},i\omega_n)\right\} \\
\lefteqn{I_{S,z}} \notag \\ 
& = & \frac{1}{4}\int_{|{\bf k}|=k_{F}}dk_{z}dk_{y}\Theta(k_{z}) \frac{k_{z}}{k_{F}}\frac{1}{\beta\hbar}\sum_{n}\sum_{\sigma}\sigma \notag \\
&&\times\left\{\frac{|\Delta_{\bf k}|}{\Omega_{n,{\bf
    k}}}a_{e\sigma\sigma}({\bf k},i\omega_n) - 
\frac{|\Delta_{-\hat{\bf k}}|}{\Omega_{n,-\hat{\bf
    k}}}a_{h\sigma\sigma}({\bf k},i\omega_n)\right\} \label{eq:spinzcur}\\
\lefteqn{I_{S,y}}\notag \\ & = & \frac{i}{4}\int_{|{\bf k}|=k_{F}}dk_{z}dk_{y}\Theta(k_{z}) \frac{k_{z}}{k_{F}}\frac{1}{\beta\hbar}\sum_{n}\sum_{\sigma}\sigma \notag \\
&&\times\left\{\frac{|\Delta_{\bf k}|}{\Omega_{n,{\bf
    k}}}a_{e\sigma\bar{\sigma}}({\bf k},i\omega_n) - 
\frac{|\Delta_{-\hat{\bf k}}|}{\Omega_{n,-\hat{\bf
    k}}}a_{h\sigma\bar{\sigma}}({\bf k},i\omega_n)\right\}
\eeqarray
where $\omega_{n} = (2n-1)\pi/\beta$, $\Omega_{n,{\bf
    k}}=\sqrt{\omega_{n}^2+|\Delta_{\bf 
    k}|^2}$ and $\bar{\sigma} = -\sigma$. 
The expression for the $x$-component of the spin current is vanishing, which
reflects the fact that
the Cooper pairs in the TSC do not have a spin component parallel to
the ${\bf d}$-vector.~\cite{Brydonspin2008} As required for the TSC-FM junction,
the charge current is found to vanish in all considered
circumstances. Furthermore, the $y$-component of the spin current is also
vanishing for the choice of ${\bf M}$ adopted here. As mentioned above, any
orientation of ${\bf M}$ can be rotated into the $x$-$y$ plane without
changing the pairing state of the TSC. Since the polarization of the spin
current  
must also be correspondingly rotated under such a transformation, we hence
conclude that the spin current is always polarized along the direction ${\bf
  d}\times{\bf M}$.~\cite{Brydonlett2009}

Due to the time-reversal-symmetry-breaking pairing state in the
$p_z+ip_y$-wave TSC, here we expect to find surface charge and
spin currents flowing parallel to the
interface.~\cite{Matsumoto1999,Kuboki2004,BrydonCFC2009} We will not be
concerned with such currents in what follows, presenting only results for the
spin current flowing perpendicular to the interface. For convenience, we shall
also adopt units where $\hbar=1$. In all plots we take the interface parameter
$Z=1$.

\section{Results}

\begin{figure}
  \includegraphics[width=\columnwidth]{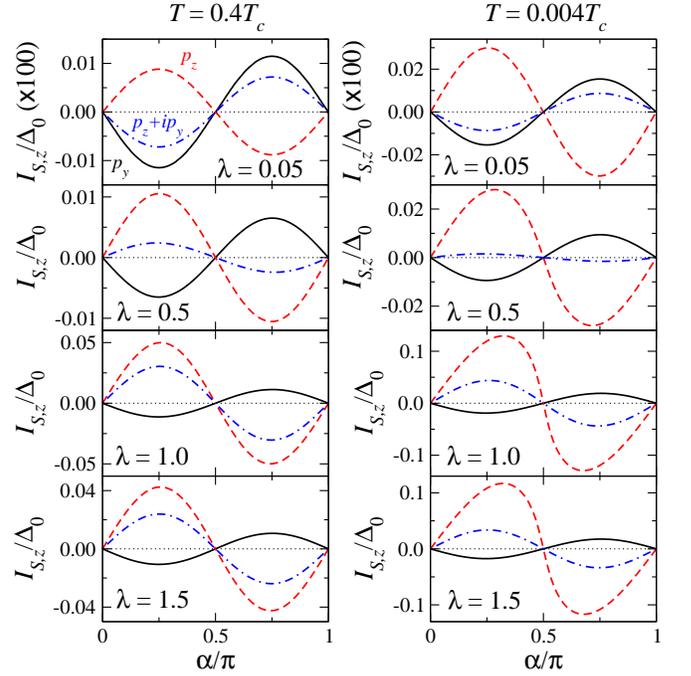}
  \caption{\label{spincurvsalpha}(color online) Spin current as a function of
  the angle $\alpha$ for the $p_y$ (black solid curve), $p_z$ (red dashed
  curve), and $p_{z}+ip_y$-wave (blue dot-dash curve) TSCs for 
  various values of the exchange-splitting $\lambda$. The left column shows
  the spin currents 
  at $T=0.4T_c$ while the right column shows the spin currents for
  $T=0.004T_c$.}
\end{figure}

We find that the spin current is an odd periodic function of $\alpha$ with
period $\pi$. As shown in~\fig{spincurvsalpha}, for most situations
the spin current obeys $I_{S,z}\propto\sin(2\alpha)$.
This dependence upon $\alpha$ is
reminiscent of the usual Josephson charge current vs phase relationship for
Josephson junctions, suggesting that
the angle of misalignment between ${\bf d}$ and ${\bf M}$ plays a role
similar to the phase difference. As was argued in~\Ref{Brydonlett2009}, this
analogy is valid: a spin $\sigma$ Cooper pair incident upon the barrier
acquires a phase shift $-2\sigma\alpha$ when undergoing a spin-flip
at reflection. Interpreting spin-flip reflection as ``tunneling''
between the spin-$\uparrow$ and spin-$\downarrow$ condensates of the TSC, the
phase shift $-2\sigma\alpha$ is therefore
the effective phase difference for a Cooper pair ``tunneling'' between the
spin $\sigma$ and spin $-\sigma$ condensates.
Naturally, this drives a Josephson charge current of
equal magnitude but opposite sign in each spin sector of the TSC, thus
producing a finite spin current but vanishing total charge current. 
Note, however, that the phase difference in a Josephson junction is a
property of the wavefunctions of the bulk condensates on either side of the
tunneling barrier; in the TSC-FM junction, in contrast, the effective ``phase
difference'' results entirely from a property of the interface.

Higher harmonics in $2\alpha$ clearly appear in the spin current for the
$p_{z}$-wave case at low temperatures and moderate to strong exchange
splitting $\lambda\gtrsim0.5$. Continuing the Josephson junction analogy, we
interpret this as evidence of coherent spin-flip reflection of multiple Cooper
pairs. In a Josephson junction, such processes are associated with the
formation of a zero energy bound state at the tunneling barrier, which allows
the resonant tunneling of multiple Cooper pairs. It is well known that
zero energy states form at unconventional superconductor interfaces when the
orbitals are aligned such that a quasiparticle specularly reflected at the
interface experiences a sign reversal of the superconducting order
parameter.~\cite{Kashiwaya2000} Such a state does not therefore occur in the
$p_y$ junction, while it is present for all
${\bf k}$ in the $p_{z}$ junction, and present only for ${\bf k}=k{\bf
  e}_z$ in the $p_z+ip_y$ junction. This is consistent with the absence of
higher harmonics in the spin current for the $p_y$ and $p_z+ip_y$
junctions. Although suggesting an
important role for the zero energy states in the spin current, it does not
explain the absence of the higher harmonics at low $\lambda$. This indicates
that the coherent reflection of multiple Cooper pairs is also controlled by the
bulk properties of the FM. 

\begin{figure}
  \includegraphics[width=\columnwidth]{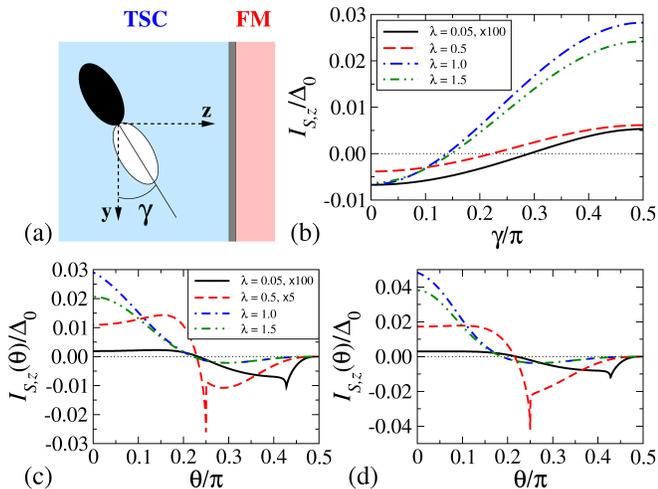}
  \caption{\label{totalrot}(color online) (a) Definition of the angle $\gamma$
    parameterizing the alignment of the $p$-wave orbital in the TSC with the
    interface. $\gamma=0$ gives the $p_{y}$-wave orbital state, while
    $\gamma=\pi/2$ gives the $p_z$-wave orbital state. (b) Spin current as a
    function of $\gamma$ for various values of $\lambda$. We take
    $\alpha=0.1\pi$ and $T=0.4T_{c}$. (c) The spin current as a
    function of the incident angle $\theta$ in the $p_{z}+ip_{y}$ junction for
    $\alpha=0.1\pi$ and $T=0.4T_{c}$; (d) shows the same curves at
    $T=0.004T_{c}$.}  
\end{figure}

It is clear from~\fig{spincurvsalpha} that the spin current is strongly
influenced by the orbital structure of the TSC; in particular, the spin
currents in the $p_{z}$ 
and $p_{y}$ junctions always have opposite sign. In the former case, the
reflected Cooper pairs acquire an additional $\pi$ phase shift due to
the reversed sign of the superconducting gap, thus reversing the spin current
relative to the $p_y$ case. 
This can be strikingly demonstrated by examining the variation of the spin
current as the $p$-wave orbital in a time-reversal-symmetric TSC is rotated
from the $p_{y}$-wave to the $p_{z}$-wave configurations. We parameterize the
orientation of the orbital by the angle $\gamma$ between the $p$-wave orbital
maximum and the $y$-axis, see~\fig{totalrot}(a). Only Cooper pairs with
incident angle $0<\theta<\gamma$ experience a reversal of the gap sign upon
reflection, and hence contribute a
spin current of opposite sign relative to the Cooper pairs with incident angle
$\gamma<\theta<\pi/2$. As shown in~\fig{totalrot}(b), the former contribution
to the spin current comes to dominate the latter as $\gamma$ is increased from
$0$ to $\pi/2$, with the spin current consequently changing sign at some
critical value of $\gamma$. 

A similar effect occurs in the $p_{z}+ip_y$ case, but here the additional
orbital phase
shift experienced by the reflected Cooper pairs is $\pi -
2\arctan(k_y/k_z)$, i.e. it depends upon the
incident trajectory. Normally incident Cooper pairs thus undergo an additional
$\pi$ phase shift as in the $p_{z}$ junction, whereas a trajectory
grazing the interface has no additional phase shift as in the $p_{y}$
junction. The angle of incidence therefore determines the sign of the spin
current contributed by the reflected Cooper pair. Making the change of
variables $k_{z}\rightarrow{k_{F}\cos(\theta)}$,
$k_{y}\rightarrow{k_{F}\sin(\theta)}$ in~\eq{eq:spinzcur}, we define the
angle-resolved spin current $I_{S,z}(\theta)$ by 
\beq
I_{S,z} = \int^{\pi/2}_{-\pi/2} d\theta I_{S,z}(\theta)
\eeq
We plot $I_{S,z}(\theta)$ in~\fig{totalrot}(c) and (d), where we see that it
changes sign at $\theta\approx0.2\pi$. Note that for $\lambda<1$ the
angle-resolved spin current is 
sharply peaked at the critical incident angle $\theta_{c}$, where the
transmitted $s=-$ quasiparticle has vanishing $z$-component of its wavevector.    

\begin{figure}
  \includegraphics[width=\columnwidth]{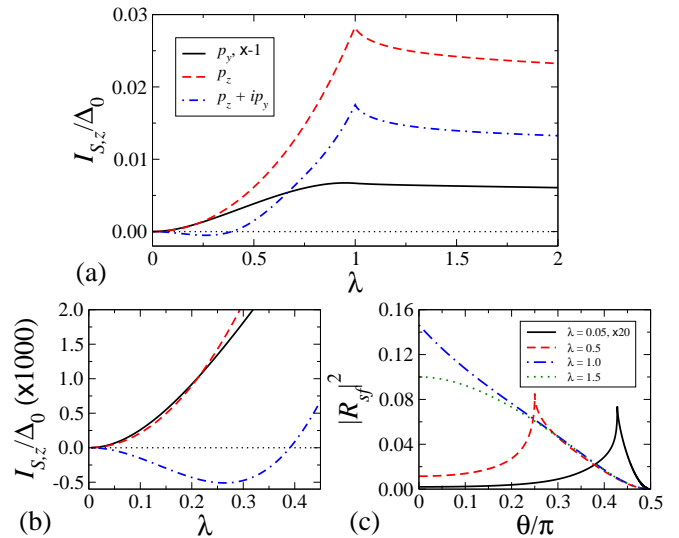}
  \caption{\label{spincurvslambda} (color online) (a) Spin current as a
    function of 
  the ratio $\lambda=g\mu_{B}M/E_{F}$. Note that the current for the
  $p_z+ip_y$-wave TSC changes sign at $\lambda\approx0.4$. (b) Spin current
  at low $\lambda$. Curves have identical meaning as in panel (a). In both (a)
  and (b) we set $\alpha=0.1\pi$ and $T=0.4T_{c}$. (c)
  $T>T_{c}$ normal state spin-slip reflection probability as a function of the
  incident quasiparticle angle $\theta$.}
\end{figure}

The variation of the spin current with $\lambda$ is shown
in~\fig{spincurvslambda}(a). There are some common features for all 
choices of the TSC orbital: at low $\lambda$ the spin current
goes as $\sim\lambda^2$; a maximum value of the spin current is reached at
($p_{z}$ and $p_z+ip_y$) or close to ($p_y$) $\lambda=1$, where the minority
spin Fermi surface disappears; and for $\lambda>1$ the magnitude of the spin
current 
displays slow monotonic decrease. Despite these similarities, the three curves
nevertheless have some key distinguishing features at $\lambda<1$. In the
$p_{y}$ case, the spin current increases much more slowly with $\lambda$ than
for the other junctions, while the maximum is smooth and occurs before the
disappearance of the minority Fermi surface. Although the spin currents in the
$p_{z}$ and $p_{z}+ip_{y}$ junctions show similar $\lambda$-dependence for
$\lambda>0.5$, the spin current in the $p_z+ip_{y}$ junction changes sign at
$\lambda\approx0.39$ [see~\fig{spincurvslambda}(b)]. We further note that the
spin current in the $p_{y}$ junction has greater magnitude than that in the
$p_{z}$ junction for $\lambda\lesssim0.2$, but the smallest magnitude of
all the currents for $\lambda\gtrsim0.65$.

The transport properties of a junction can be
formulated in terms of the normal state scattering matrix.~\cite{spinactivebcs}
Insight into the complicated dependence of the spin current upon $\lambda$ can
therefore be gained by examining the $T>T_{c}$ spin-flip reflection
probability $|R_{sf}|^2$, given by
\beq
|R_{sf}|^2 = \begin{cases} 
\frac{k_{z}^2\left(k_{+,z}-k_{-,z}\right)^2}{\left[Z^2k_{F}^2 +
    (k_{z} + k_{-,z})^2\right]\left[Z^2k_{F}^2 +
    (k_{z} + k_{+,z})^2\right]} & |\theta| < \theta_c \\
\frac{k_{z}^2\left(k_{+,z}^2+\kappa^2\right)}{\left[k_{z}^2 +
    (Zk_{F} + \kappa)^2\right]\left[Z^2k_{F}^2 +
    (k_{z} + k_{+,z})^2\right]} & |\theta| \geq \theta_c
\end{cases}
\eeq
Note that $|R_{sf}|^2$ is finite when $Z=0$, and so the spin current survives
when the insulating barrier is removed. We plot $|R_{sf}|^2$ for $Z=1$
in~\fig{spincurvslambda}(c). When $\lambda\leq1$ the
spin-flip reflection probability is sharply peaked at the critical angle
$\theta_{c}$, and grows in magnitude as $\lambda$ is increased. For
$\lambda>1$ the reflection 
probability remains peaked at $\theta=0$, although it decreases from the
$\lambda=1$ maximum. For larger values of $Z$, $|R_{sf}|^2$ may very slightly
increase with increasing $\lambda$ at $\theta\neq0$ (not shown). Comparison of
the angle-resolved spin current $I_{S,z}(\theta)$ 
in~\fig{totalrot}(c) and (d) with~\fig{spincurvslambda}(c) clearly shows the
connection of $|R_{sf}|^2$ to the transport: the spin-flip reflection of
Cooper pairs with incident angle $\theta$ close to $\theta_{c}$ is strongly
favored, and tends to dominate the spin current. The sign change of
the spin current in the $p_{z}+ip_{y}$ junction with increasing $\lambda$ can
thus be understood as arising from the shift in the peak of $|R_{sf}|^2$: at
$\lambda\ll1$ this strongly favors grazing trajectories experiencing only small
orbital phase shift, while at $\lambda\sim1$ almost-normal trajectories with
orbital phase shift close to $\pi$ dominate.

The gap anisotropy in the $p_{z}$ and $p_y$ junctions 
implies a greater contribution to the spin current for trajectories along which
the gap magnitude is maximal. The peak in $|R_{sf}|^2$ at grazing trajectories
for $\lambda\ll1$ therefore results in a greater spin current in the $p_y$
junction than the $p_z$ junction. As $\lambda$ is increased, the reduction of
$\theta_{c}$ leads to the observed strong enhancement of the spin current in
the $p_z$ junction. Because the reflection
probability at high $\theta$ also grows with $\lambda$, the spin current
continues to increase in the $p_y$ junction, albeit at a much slower rate than
in the $p_z$ junction. The reduction of the spin current at $\lambda>1$ is
due to the reduction of the peak in $|R_{sf}|^2$ at
$\theta=0$. 

\begin{figure}
  \includegraphics[width=\columnwidth]{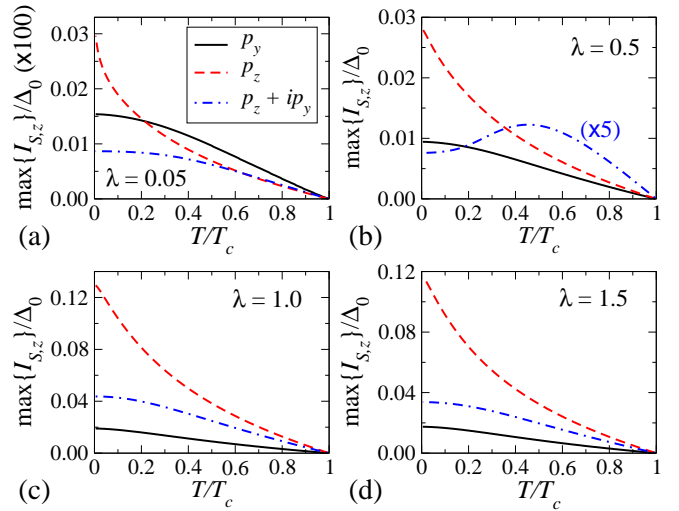}
  \caption{\label{spincurvsmaxT} (color online) The maximum value of the spin
    current obtained by varying $\alpha$ for (a) $\lambda=0.05$, (b)
    $\lambda=0.5$, (c) $\lambda=1.0$ and (d) $\lambda=1.5$. The curves in
    panels (b)-(d) have the same meaning as those in panel (a); note that the
    $p_{z}+ip_{y}$ curve is multiplied by 5 in (b) for clarity.}
\end{figure}

The last quantity of interest is the maximum spin current $\max\{I_{S,z}\}$
obtained by varying $\alpha$. This is plotted as a function of $T$ for fixed
$\lambda$ in~\fig{spincurvsmaxT}. As 
can be seen, for all $\lambda$ the magnitude of the spin current in the $p_y$
junction shows a moderate enhancement with decreasing temperature, with an
apparent plateauing at $T<0.1T_{c}$. In the
$p_{z}$ junction, in contrast, there is a strong enhancement 
of the spin current magnitude with decreasing temperature. The temperature
dependence of $\max\{I_{S,z}\}$ in
the $p_z$ and $p_y$ cases is reminiscent of the critical current in Josephson
junctions with and without zero energy states
respectively,~\cite{Kashiwaya2000,Kwon2004} again indicating a role for such
states in the device studied here. The temperature dependence of
$\max\{I_{S,z}\}$ in the $p_{z}+ip_{y}$ junction is determined by $\lambda$:
at low and high $\lambda$ we find the moderate increase with decreasing $T$
characteristic of the $p_{y}$ junction, and also seen in the critical current
of non-magnetic
Josephson junctions between $p_{z}+ip_{y}$ TSCs.~\cite{Barash2001} At
intermediate $\lambda$, however, we find a non-monotonic dependence of
$\max\{I_{S,z}\}$ upon $T$. This can be understood by examining the
angle-resolved spin current in~\fig{totalrot}(c) and (d): for $\lambda=0.5$,
we observe a particularly strong enhancement of the current for trajectories
near $\theta_{c}$ as the temperature is lowered. Since the current for such
trajectories has opposite sign to 
that of the total current, there is consequently a reduction of the total spin
current. 
For a small range of $\lambda\sim0.4$, the spin current in
the $p_{z}+ip_{y}$ junction reverses sign with decreasing temperature.

\section{Conclusions}

This work presents an analysis of the spontaneous spin current generated in a
TSC by contact with a bulk metallic FM. The
essential requirement for the appearance of this spin current 
is that the vector order parameters of the TSC and FM are misaligned but not
mutually perpendicular. Following~\Ref{Brydonlett2009}, the spin current has
been interpreted as a Josephson-like effect, where the phase shift picked up
by Cooper pairs undergoing spin-flip reflection at the interface plays the
role of the phase difference between the superconductors in a Josephson
junction. This analogy is supported by the temperature evolution of the
maximum spin current, and the low-temperature dependence upon $\alpha$, which
suggests that resonant reflection of multiple Cooper pairs through zero energy
states plays a significant role in the current.

The spin current nevertheless possesses several properties which are not
anticipated by the Josephson junction analogy. As it arises from reflection
processes, the choice of orbital pairing state in the TSC can determine the
sign of the spin current, due to the orbital phase shift experienced by the
reflected Cooper pairs. In the case of the $p_{z}+ip_{y}$ junction, this
produces the interesting result that spin-flip reflected Cooper pairs with
different incident trajectories can carry spin currents of opposite sign. 
The exchange-splitting of the FM also influences the spin
current, most significantly when the FM has both a minority and majority spin
Fermi surface. This is closely connected to the orbital structure of the TSC,
as the exchange-splitting determines the angular-dependence of the spin-flip
reflection probability. 

\begin{figure}
  \includegraphics[width=0.5\columnwidth]{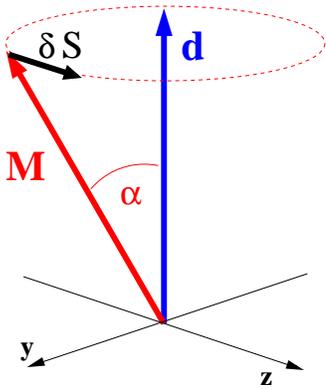}
  \caption{\label{precession} (color online) The spin-flip reflection of the
    Cooper pairs imparts the spin change $\delta{\bf S} \parallel {\bf
      d}\times{\bf M}$ to the FM. We speculate that this results in a
    precession of the magnetic moment around the axis defined by the ${\bf d}$ 
    vector.}  
\end{figure}

A few critical remarks upon our method are necessary. 
Considering that interface effects play the essential role in the generation
of the spin current, it is reasonable to question our approximation that the
TSC and FM order parameters are both constant up to the insulating
barrier. It is well known that there is a strong suppression of the
$p_{z}$-wave gap at partially-transparent
interfaces,~\cite{Matsumoto1999,Tanuma2006} but as this does not change the
spin structure of the TSC it is only likely to
renormalize the results presented here. More interesting is a possible change
in the orientation of ${\bf d}({\bf r})$ and ${\bf M}({\bf r})$ close to the
interface due to an 
unconventional proximity effect. Although the analysis of~\Ref{Grein2009}
indicates that the spin current will 
survive a self-consistent treatment of the superconductor, it is 
desirable that the TSC and FM be treated on an equal level. This is possible,
for example,  
within a real-space Hartree-Fock analysis.~\cite{Zhu2000,Asano2004,Kuboki2004,Cuoco2008} The TSC-FM interface has already
been studied using this method,~\cite{Kuboki2004,Cuoco2008} but only for the
case ${\bf d}\parallel{\bf 
  M}$ which we predict to display vanishing spin current. Much scope therefore
remains for further investigations.

Another important assumption is that the magnetic moment of the FM is
constant. As pointed out in~\Ref{Brydonlett2009}, however, the spin-flip
reflection of the Cooper pair imparts a spin $\delta{\bf S} \parallel {\bf
  d}\times{\bf M}$ to the FM. If $|\delta{\bf S}|\ll|{\bf
  M}|$, the effect of the imparted spin should be to slightly rotate the
magnetic moment about the axis defined by ${\bf d}$, without significantly
altering its magnitude, see the cartoon representation
in~\fig{precession}. This is a generalization of the well-known spin-transfer
torque effect of spintronics to a superconducting device.~\cite{Waintal2000}
For a FM without anisotropy, we 
speculate that the cumulative effect of many spin-flip reflection events is
to cause the precession of the magnetic moment. If this precession is
sufficiently slow, the results presented here should still remain valid, and
we anticipate a periodic modulation of the polarization of the spin current
due to the rotating moment. Rigorous verification of these claims requires a
more sophisticated analysis than our argument above, and is left for future
work. 

Finally, we consider the prospects for the experimental verification of our
predictions. Direct measurement of spin currents in superconductors is
unlikely to be easy, although several proposals exist, e.g. spin-resolved
neutron scattering~\cite{Hirsch1999} or ARPES with circularly-polarized
light.~\cite{Simon2002} Alternatively, one could search for evidence of a
spin-accumulation at the edge of the TSC opposite to the interface with the
FM. If our speculation above proves to be well-founded, it
might also prove possible to deduce the existence of a spin current by measuring
the precession of ${\bf M}$. Initial characterization of a TSC-FM device
would, however, almost certainly involve measurements of the LDOS at the
interface~\cite{Yokoyama2007b} and the tunneling conductance.~\cite{Hirai}
The unique behaviour of these quantities at TSC-FM interfaces is a signature
of the unconventional interplay of ferromagnetism and triplet
superconductivity, of which the spontaneous spin current is only one
manifestation. 

\acknowledgments

The author thanks R. Grein and M. Eschrig for discussions on their
work~\Ref{Grein2009}. B. L. Gy\"{o}rffy and Y. Tanaka are acknowledged for
pointing out several important references. Special thanks is given to
D. Manske and C. Timm for their helpful advice and careful reading of the
manuscript.


\begin{thebibliography}{99}
\bibitem{simplify}Relaxing these simplifying assumptions does not
  qualitatively alter our results.
\end{thebibliography}
\end{document}